\documentclass[%
 reprint,
 floatfix,
 superscriptaddress,
 amsmath,
 amssymb,
 longbibliography
]{revtex4-1}

\usepackage{graphicx}% Include figure files
\usepackage{dcolumn}% Align table columns on decimal point
\usepackage{bm}% bold math
\usepackage{tikz}
\usepackage{braket}
\usepackage{amsmath}
\usepackage{appendix}
\usepackage{xcolor}
\usepackage{float}
\usepackage{tikz}
\usepackage[utf8]{inputenc}
\DeclareUnicodeCharacter{00B2}{\ensuremath{{}^2}}

\usepackage{blochsphere}
\usepackage[caption=false]{subfig}
\usepackage{algorithm2e}

\newcommand{\mEh}{{mE$_{\mathrm{h}}$}}
\newcommand{\nmom}{{$n_{\mathrm{mom}}$}}
\newcommand{\Omeas}{{${\hat O}_{{\mathrm{mes}}}$}}
\newcommand{\Oerr}{{${\hat O}_{{\mathrm{err}}}$}}

\begin{document}

\preprint{APS/123-QED}

\title{Reduced Density Matrix Sampling: Self-consistent Embedding and Multiscale Electronic Structure on Current Generation Quantum Computers}

\author{Jules Tilly}
 \email{jules.tilly@rahko.ai}
 \affiliation{Rahko  Limited,  N4 3JP London,  United  Kingdom}
 \affiliation{Department of Physics and Astronomy, University College London, WC1E 6BT London, United  Kingdom}
\author{P.V. Sriluckshmy}
 \affiliation{Department of Physics, King's College London, Strand, London, WC2R 2LS, U.K.} 
\author{Akashkumar Patel}
 \affiliation{National Physical Laboratory, TW11 0LW Teddington, United Kingdom}
 \affiliation{Department of Physics and Astronomy, University College London, WC1E 6BT London, United  Kingdom} 
\author{Enrico Fontana}
 \affiliation{National Physical Laboratory, TW11 0LW Teddington, United Kingdom}
 \affiliation{Department of Computing, University of Strathclyde, G1 1XQ Glasgow, United Kingdom}
\author{Ivan Rungger}
 \affiliation{National Physical Laboratory, TW11 0LW Teddington, United Kingdom}
  \affiliation{Department of Physics and Astronomy, University College London, WC1E 6BT London, United  Kingdom}
 
\author{Edward Grant}
\affiliation{Department of Computer Science, University College London, London WC1E 6BT, UK}
\affiliation{Rahko  Limited,  N4 3JP London,  United  Kingdom}
\author{Robert Anderson}
\affiliation{Department of Physics, King's College London, Strand, London, WC2R 2LS, U.K.}
\author{Jonathan Tennyson}
 \affiliation{Department of Physics and Astronomy, University College London, WC1E 6BT London, United  Kingdom}
 \author{George~H.~Booth}
\affiliation{Department of Physics, King's College London, Strand, London, WC2R 2LS, U.K.}
\date{\today}% It is always \today,  today,
             %  but any date may be explicitly specified

\begin{abstract}
We investigate fully self-consistent multiscale quantum-classical algorithms on current generation superconducting quantum computers, in a unified approach to tackle the correlated electronic structure of large systems in both quantum chemistry and condensed matter physics. In both of these contexts, a strongly correlated quantum region of the extended system is isolated and self-consistently coupled to its environment via the sampling of reduced density matrices. We analyze the viability of current generation quantum devices to provide the required fidelity of these objects for a robust and efficient optimization of this subspace. We show that with a simple error mitigation strategy and optimization of compact tensor product bases to minimize the number of terms to sample, these self-consistent algorithms are indeed highly robust, even in the presence of significant noises on quantum hardware. Furthermore, we demonstrate the use of these density matrices for the sampling of non-energetic properties, including dipole moments and Fermi liquid parameters in condensed phase systems, achieving a reliable accuracy with sparse sampling. It appears that uncertainties derived from the iterative optimization of these subspaces is smaller than variances in the energy for a single subspace optimization with current quantum hardware. This boosts the prospect for routine self-consistency to improve the choice of correlated subspaces in hybrid quantum-classical approaches to electronic structure for large systems in this multiscale fashion.

\end{abstract}
\maketitle

\section{Introduction}
A solution to the quantum many-body problem is held up as a one of the most impactful and far-reaching applications of quantum computers \cite{Reiher2017,doi:10.1021/acs.jctc.9b00236,Mccaskey19,bauer2020quantum}. Even in the current era of noisy intermediate-scale quantum (NISQ) devices \cite{Preskill2018}, where the number of physical qubits is too small for error correction and subject to significant decoherence and quantum noise, significant progress in this area has been made, with developments also impacting the related field of quantum machine learning \cite{xia18}. However, quantum resources are finite, and so treating entire systems of technological relevance is unlikely to be a realistic near-term proposition. Instead, efficient and practical hybrid quantum-classical and multi-scale approaches are key to describing quantum systems with a desired accuracy within a limited quantum computational budget.

The general aim of these is to find a representation of the full system in which only a small number of `active' degrees of freedom are strongly entangled and require realization on a quantum device. The remaining degrees of freedom are then coupled to this quantum subsystem in a low-rank perturbative or even (potentially dynamical) mean-field representation of the quantum effects. These `external' weakly-interacting degrees of freedom can be efficiently described on a classical computer with polynomially-scaling resources with system size, while it is posited that an appropriate choice of strongly entangled `active' orbitals should not in general need to grow with system size. Self-consistency can then be used to update the initial choice of the active region, based on a multi-level description of the full system. 
One of the key technical considerations in devising these approaches is in the choice of how the description of the active space on the Quantum Processing Unit (QPU) can be efficiently coupled to the external space. In this work, we consider the sampling of reduced density matrices (RDMs) as the natural choice of low-rank quantum variable for NISQ computers, in order to describe the coupling of the classical and quantum realizations of the system and therefore straddle the length scales in the simulation of quantum matter. We present a unified description of this approach, with applications to both strongly correlated quantum chemistry and condensed matter problems on current generation quantum devices.

While multi-level descriptions of quantum systems are in themselves not a new proposition for quantum computers \cite{Bauer2015,GalliQuantumEmbedding,yamazaki2018practical,2016arXiv161006910R,dhawan2021dynamical,Takeshita2020,Yao2020,rossmannek2020quantum, Yalouz_2021}, the practical realization of fully self-consistent algorithms on quantum resources has proved to be a significant technical challenge. The ability of  self-consistent coupled quantum-classical approaches to remain robust in the presence of the noise inherent in the sampling of the active space quantum effects is a key practical consideration. We demonstrate that with a light-touch error mitigation strategy, noise does not preclude stable convergence of the algorithms presented in this paper (although the variance of the final energetics can remain high). Furthermore, an efficient sampling and grouping scheme for the terms required in density matrices is presented, detailing a huge reduction in the number of these terms for a given active space size, which tackles a key bottleneck for longer-term practical quantum-classical multiscale methods.

We demonstrate the feasibility of two of these self-consistent algorithms in a unified approach applicable to both quantum chemistry and condensed matter physics. The first is complete active space self-consistent field (CASSCF) \cite{Roos1980, Sun2017}, a powerful approach for the simulation of molecular systems with strong quantum effects, such as those encountered routinely in inorganic chemistry, systems with competing spin states, excited states, and systems at bond-breaking geometries \cite{RetaManeru2014,Li2015,olsen11}. 
In these problems, the dominant strong quantum fluctuations can often be qualitatively captured within a small number of low-energy orbitals, where these orbitals are obtained from a prior mean-field calculation. However, since this starting point does not account for the correlated physics, these active orbitals are subsequently optimized in a variational fashion across the overall state as a product of the active space wave function and mean-field electrons in the external space. 

As systems get larger, and the spectrum of single-particle states transitions towards a continuous function of energy, the approach of choosing quantum regions using an energetic criteria becomes less well defined. In this case, it is more appropriate to consider the active quantum region in a real-space local picture, under the assumption that local quantum fluctuations dominate, such as those in the $d$-orbitals of a transition metal oxide material. This is the approach taken in various quantum embedding methods such as dynamical mean-field theory \cite{Maier2005}, and in this work we consider the related energy-weighted density matrix embedding theory (EwDMET) \cite{Fertitta2018,Fertitta2019,Sriluckshmy2021}. We perform the self-consistency at the level of energy weighted density matrices, denoting the moments of the local density of states, resulting in a systematic expansion of the zero-temperature DMFT physics \cite{Sriluckshmy2021}. For both of these approaches, we demonstrate the fidelity of the QPU sampling of the active space RDMs required for a fully QPU-coupled self-consistent algorithm, and consider the scaling of sampling operations as the active space increases in size in future applications.

In section~\ref{sec:RDM} we review reduced density matrices and their sampling within QPUs as the self-consistent quantum variables in these multi-scale methods. We demonstrate that judicious grouping of commuting terms allows even large active space RDMs to be realistically sampled, with the proposed groupings opening the prospect for higher-rank RDMs and perturbative couplings to active spaces. In section~\ref{sec:QCASSCF} we consider the CASSCF optimization of carbon monoxide on IBM Quantum services (IBMQ) machines, where active space wave function are optimized on the QPU within the variational quantum eigensolver (VQE) \cite{Peruzzo2013}. We find a naive implementation to preclude full self-consistency, however a light-touch error mitigation strategy is enough to address this issue on a 4-qubit system and to allow convergence of the active space orbitals in the presence of the correlated physics. The sampling noise of the QPU does not prevent convergence, and the CASSCF is found to be robust and reliable in its optimization. We also stress that knowledge of the active space RDMs also allows for the extraction of beyond-energetic first-order expectation values of the system, which are essential for a more complete description of the system properties, focusing on the effect of self-consistency and RDM fidelity on the dipole moment of a system.

Finally, in section~\ref{sec:QEwDMET} we focus on extended bulk systems, with the strongly correlated Bethe-Hubbard lattice considered. Specifically, we observe the QPU description of a local region to allow for the opening of Hubbard bands in the material within the QPU-coupled EwDMET approach. Similar error mitigation strategies on the sampled active space RDMs allows for robust self-consistency in the method, resulting in excellent agreement for the local density of states and Matsubara self-energy of the system on current generation IBMQ machines.

\section{Sampling Reduced Density Matrices on a Quantum Computer} \label{sec:RDM}

The reduced density matrices (RDM) used in this work are not defined by tracing out a subsystem, but rather tracing out the entire phase space of many electrons from the full $N$-particle density matrix of a pure state. For a $m$-body reduced density matrix, $\Gamma_m$, this integration over $N-m$ electronic variables can be written as
\begin{align}
    &\Gamma_m({\bf x}_1, \dots ,{\bf x}_m ; {\bf x}'_1, \dots , {\bf x}'_m) = m!  \binom{N}{m}  \times \\
    &\int \Gamma_N(\{{\bf x}\}_N ; \{{\bf x}'\}_N) \Delta_{m+1}^N d{\bf x}_{m+1} \dots d{\bf x}_N d{\bf x}'_{m+1} d{\bf x}'_N, \nonumber
\end{align}
where ${\bf x}_i$ represents the combined spatial and spin coordinate for electron $i$, and $\Delta_{m+1}^N=\prod_{i=m+1}^N \delta({\bf x}_i - {\bf x}'_i)$. Fortunately, these reduced-body density matrices can be directly computed, rather than requiring tracing from higher-rank density matrices. By projecting the electronic coordinates into a basis set, we define the two-body RDM as
\begin{equation}
    \Gamma_{ijkl}
    \equiv
    \bra{\psi} \hat{a}_i^{\dagger}\hat{a}_j^{\dagger} \hat{a}_l \hat{a}_k \ket{\psi},
    \label{eq:2}
\end{equation}
with other rank RDMs defined equivalently, and 
where the indices $i, j, \dots$ label spin-orbital degrees of freedom, ${\hat a}_i^{(\dagger)}$ are the fermionic annihilation (creation) operators, and we have omitted the explicit subscript denoting the rank of the RDM where it is obvious from the number of indices. In this example, the partial trace down to the one-body RDM can then be written as
\begin{equation}
    \gamma_{ik} = \frac{1}{N-1}\sum_j\Gamma_{ij,kj}.
\end{equation}
Despite tracing out large numbers of degrees of freedom, these two-body RDMs still contain all the information about a quantum system required for physical observables of interest which depend on (up to) pairwise operators, including the total energy. The rank of an operator defining a given observable determines the rank of the RDM required to compute its corresponding expectation value. For example, the electric dipole moment is a one-body quantity, requiring the one-body RDM, while the Hamiltonian defining the energy is a two-body expectation value, requiring the two-body RDM to evaluate. 
Non-observable quantities of interest, such as entanglement entropies or mutual information, can also in general be computed from reduced-body density matrices \cite{doi:10.1021/ct400247p}.

Furthermore, using RDMs we can compute the probability of a given $m$-electron distribution, as the diagonal of the $m$-RDM. The sum over this distribution then gives the number of $m$-tuples of particles in the system, which can be used as a normalization condition, e.g. the trace of the 2-electron distribution giving the number of pairs of electrons, as
\begin{equation}
    \sum_{ij}\Gamma_{ij,ij} = \frac{N(N-1)}{2}.
\end{equation}
Overall, these $m$-RDMs have all the information about the distribution and entanglement of $m$ particles in a given state of an $N$ particle system, which rationalizes their use as method-agnostic, low-rank quantities in order to couple quantum systems described at different levels of theory. 

In this work, we consider second-quantized Hamiltonians where spin symmetry is preserved, allowing further tracing out of spin degrees of freedom, defining the central spin-free two-body RDM of interest as
\begin{equation}
    \overline{\Gamma}_{pqrs} \equiv \sum_{\sigma \tau} \Gamma_{p_\sigma q_\tau r_\sigma s_\tau}, \label{eq:spin2RDM}
\end{equation}
where $p, q,\dots$ denote spatial degrees of freedom and $\sigma, \tau$ denote spin labels. Further permutational symmetries can be used which reduce the number of independent quantities to evaluate, as
\begin{equation}
    \Gamma_{p_\sigma q_\tau r_\sigma s_\tau} = -\Gamma_{p_\sigma q_\tau  s_\tau r_\sigma} = 
    \Gamma_{q_\tau p_\sigma s_\tau r_\sigma} = \Gamma_{r_\sigma s_\tau p_\sigma q_\tau}, \label{eq:antisym2RDM}
\end{equation}
with time-reversal symmetry ensuring
\begin{equation}
    \Gamma_{p_\sigma q_\tau r_\sigma s_\tau} = \Gamma_{p_{\bar{\tau}} q_{\bar{\sigma}} r_{\bar{\tau}} s_{\bar{\sigma}}}. \label{eq:time-rev2RDM}
\end{equation}

The resulting set of fermionic operators must be mapped to spin operators for sampling of the state on a QPU. For this, we use the Jordan-Wigner mapping \cite{Jordan1928}, though other mappings (e.g. Bravyi-Kitaev mapping \cite{Bravyi2002, Seeley2012}) could be used, as long as the mapping is consistent. 
While each fermionic operator will in general be mapped to several spin operators, one can find several efficiencies to reduce this overall number of terms. 

At most, sixteen Pauli strings result from each two-body fermionic operator. However these strings resulting from different fermionic operators often commute, and therefore do not require separate measurement. Once all unique terms have been isolated, they can be subsequently grouped into a Tensor Product Basis (TPB) to reduce the computational cost further (for a recent review of grouping strategies, see Ref.~\onlinecite{Hamamura2020}).
These grouping strategies tend to be computationally expensive, however the final set of non-commuting Pauli strings required to measure a given rank of RDM will be identical and agnostic to the details of the Hamiltonian for a given number of orbitals.

As a result, once an optimal set of terms is established, it can be used for all systems, in a similar fashion to the energy measurement for a VQE problem of a given size (which is equivalent to the two-body RDM).  \\

\begin{figure}
    \centering
    \includegraphics[width=\linewidth]{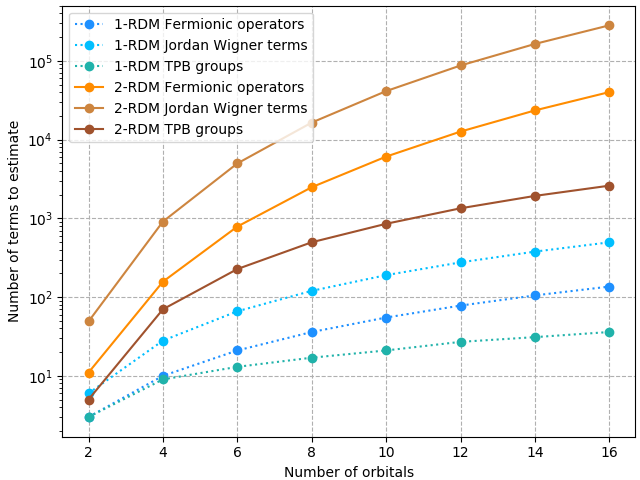}
    \caption{Number of unique fermionic operators, corresponding set of unique Pauli strings (under Jordan-Wigner mapping), and Tensor Product Basis groups to be measured in order to compute all elements of the one-body (dotted line) and two-body RDM (solid line), for up to 16 orbitals (32 qubits). The Tensor Product Basis was found via the Largest-degree First Coloring (LDFC) algorithm detailed in Appendix~\ref{sec:RDM_group_tables}.}
    \label{fig:grouping_numbers}
\end{figure}

We present in Fig. \ref{fig:grouping_numbers} the number of terms in the TPB for the one- and two-body RDMs as a function of the number of molecular orbitals in the active space (where each molecular orbital is mapped to two qubits). Our results are similar to those found previously in the literature (see for instance Ref.~\cite{vallury_2020}), showing a significant reduction in dimensionality and scaling of the TPB grouping of terms as the active space increases in size. This results in a reduction of over two orders of magnitude for the number of terms to be sampled in a 16 orbital active space, with this factor  increasing for larger active spaces. We discuss the method used to obtain the TPB groups as well as details of the overall number of terms to sample in Appendix~\ref{sec:RDM_group_tables}, while also presenting similar groupings for three- and four-body RDMs, required for many multireference perturbative or subspace expansion approaches going forwards to ameliorate the restriction on active space sizes \cite{Angeli2007,Takeshita2020,doi:10.1063/1.5140086}.
One point of note is that the symmetries used in Eqs.~\ref{eq:spin2RDM}-\ref{eq:time-rev2RDM}, combined with the Jordan-Wigner mapping, ensure that the one-body RDM only relies on half of the wave function (the same half of all Pauli strings required are identities, rendering half of the qubits used obsolete in the sampling). This feature can be used to easily sample the energy of any product state or one-body RDM functional \cite{QuantumHF}.

It is worth noting that TPB grouping of terms may entail additional costs. Firstly, the joint measurements of Pauli operators results in a covariance between terms, potentially increasing the overall variance of the observable expectation values. In exceptional cases, this can even increase the total number of samples required for a given fidelity \cite{McClean_2016}. In general however, we should expect a reduction in number of measurements necessary for a given precision \cite{Gokhale_2019}. Secondly, joint measurements of TPB groups require additional circuit depth to rotate the measurement basis appropriately. This additional circuit scales $O(N^2)$, with $N$ the number of qubits \cite{Gokhale_2019}, and therefore should be considered small on larger gate depth circuits (by comparison, the Generalized Unitary Coupled Cluster Ansatz scales $O(N^3)$ in depth. See for instance Ref. \onlinecite{Lee_2019}). In the case of current generation QPUs however, this additional circuit length (largely composed of entangling gates) results in quantum noise that would arguably out-weigh the benefits obtained from reduction of finite sampling noise from operator joint measurements. For this reason, while this reduction in terms to sample is a promising feature for longer-term viability of density matrix sampling, we are unlikely to benefit from this for small qubit arrays on current generation QPUs, and therefore leave the use of operator grouping in actual experiments for future work.

\begin{figure}
    \centering
    \includegraphics[width=\linewidth]{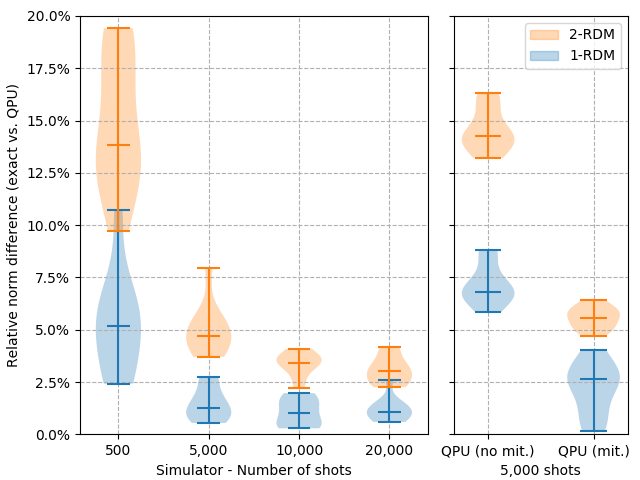}
    \caption{Distribution of relative errors in the Frobenius norm for both QPU and quantum emulated sampling of the RDMs, compared to exact classical calculation. To obtain the distributions, we repeated the computation of the Frobenius norms differences over 20 realizations of the RDMs for each number of measurements considered. On the left, results are presented for a QPU simulator (assuming perfect qubits) and therefore displays the impact of finite sampling noise. On the right, the results computed on IBMQ Athens (4-qubits, depth 3 HEA), with and without error mitigation.}
    \label{fig:rdm_plot}
\end{figure}

In order to test RDM sampling on a quantum computer, we computed the one- and two-body RDMs of magnesium porphyrin after optimization of a hardware efficient ansatz (HEA) wave function, with an active space restricted to 2 orbitals and 2 electrons (a 4 qubit Hamiltonian). The resulting RDMs are then compared to the RDMs obtained via exact methods for the same active space, and the distribution of relative errors in the elements are shown in Fig.~\ref{fig:rdm_plot}. We conducted this test on a simulated QPU with different number of shots, as well as current quantum hardware (IBMQ Athens QPU, details about each QPU used in this paper can be found in Appendix \ref{sec:lattices})). \\

In our investigation of RDM observable sampling errors, we also considered approaches to reliably mitigate for these errors via physically justified extrapolation techniques. It is clear from the results of Fig.~\ref{fig:rdm_plot} that these can substantially ameliorate quantum noise and sampling errors. We present the simple extrapolation technique used in this work in Appendix~\ref{sec:error_mitigation}; the technique relies on a binomial distribution of independent errors in the quantum circuit, and has the benefit of requiring no additional measurements (for a more involved approach to extrapolation, we recommend Ref.~\onlinecite{Endo2018_PRX}).

This simple error mitigation technique significantly improves the overall accuracy of the QPU estimates, bringing it almost on par with the results of the simulator at an equivalent number of measurements. A key question that remains is whether the norm error presented above has a significant impact on the ability to use these RDMs reliably within subsequent quantum chemical calculations, where manipulation of these noisy RDMs may prevent convergence or lead to unacceptable bias in desired quantities.
In order to test this we apply this sampling to a QPU-solved complete active space self-consistent field method (computing both energetics and molecular dipole moments following the optimization), as well as a QPU version of the energy-weighted density matrix embedding theory, as examples of multi-scale approaches to allow quantum resources to be applied to realistic systems in electronic structure calculations.

\section{Quantum CASSCF} \label{sec:QCASSCF}

The complete active space self-consistent field (CASSCF) approach is generally the starting point in quantum chemistry for molecular systems exhibiting stronger correlation effects, and therefore a key step in the development of electronic structure methods suitable for quantum computation \cite{Lengsfield1980, Roos1980, Werner1980, Werner1985, Werner1987, Olsen1983, Jensen1984, Shepard1987, Tenno1996, Schmidt1998, Nakano2000, Aquilante2008, olsen11, Szalay2012, Gyorffy2013, Kim2015, Hohenstein2015}. 
The central tenet of CASSCF is that the dominant strong quantum fluctuations required to qualitatively describe an electronic system are spanned by a small number of low-energy degrees of freedom about the chemical potential. The changes caused by explicitly considering interaction-driven virtual excitations in this space can change the occupation and induce entanglement of these orbitals, giving rise to correlated physics far from a mean-field description. 
The first step of CASSCF is therefore to partition the orbitals into three subspaces, denoted core, active and virtual. Core orbitals are deep-lying orbitals, which are considered to be chemically-inert and fully occupied, while conversely, the virtual orbitals are considered high-energy states which remain unoccupied. Together, these denote the `external' space. The active space denotes the degrees of freedom which are considered to span the dominant electron correlations corresponding to low-energy virtual excitations of the $N_{\mathrm{act}}$ electrons within it, with the full set of quantum fluctuations amongst this set to be considered. No entanglement or particle/spin fluctuations are considered between the external and active spaces. The overall CASSCF wave function at any point can therefore be written as
\begin{equation}
    |\Psi_{\mathrm{CASSCF}} \rangle = |\psi_{\textrm{active}} \rangle \otimes \textrm{det}[\phi_c] , \label{eq:CASSCFWfn}
\end{equation}
where $|\psi_{\textrm{active}} \rangle$ denotes an $N_\textrm{act}$-electron wave function spanning the active degrees of freedom, while $\textrm{det}[\phi_c]$ is a single product state over the core orbitals, accounting for the $N-N_{\mathrm{act}}$ remaining electrons.

A key initial step for CASSCF is therefore to choose the orbitals in each set. These are selected from an initial mean-field calculation, where to a first approximation, the highest-energy occupied and lowest-energy virtual orbitals about the chemical potential are chosen as the active space. However, this choice is often augmented with other criteria for selection of the active space, including symmetry, locality and/or `chemical intuition', with approaches for automatic selection of this space e.g. from quantum information arguments, a source of recent developments \cite{Reiher2017,Gagliardi2020, Sayfutyarova2017}. However, it is clear that selecting these orbitals from an initial mean-field calculation has an inherent flaw. The active space, designed to capture the strong correlations and dominant entanglement between single-particle states, is chosen from a theory with no correlations or entanglement via simple mean-field orbital energetics, which can change substantially in the presence of electron correlation. To account for this, a self-consistency in the choice of the active space orbitals is required for meaningful and qualitatively accurate results in the presence of strong correlation. This involves a variational optimization of the state given in Eq.~\ref{eq:CASSCFWfn}, to account for an arbitrary mixing between all three classes of orbitals, defined by the exponential of an anti-hermitian one-particle operator. This allows the character of the orbitals to change, by rotating core and virtual components into the active space in a variationally optimal way.

The CASSCF method from another perspective can be considered as an embedding of the correlated effects of the active space into a mean-field description of its `environment' (as given by the electrons in core orbitals), as presented in Ref.~\onlinecite{rossmannek2020quantum} for quantum emulation. However, this active `embedding region' is chosen largely on energetic criteria, with a strictly separable form and no entanglement with the core electrons. We contrast this with an alternative criteria based on locality in Sec.~\ref{sec:QEwDMET}. The limitations of the approach come from the size of the active space, which for an exact treatment is often accepted to be 16 electrons in 16 orbitals \cite{olsen11}, with some instances of computation up to 20 electrons in 20 orbitals \cite{Vogiatzis2017}. This is due to the factorial scaling of classical resources with respect to this size in order to represent and optimize $|\psi_{\textrm{active}} \rangle$. Beyond this, approximate descriptions of the active space wave function are increasingly being investigated, although all have their limitations \cite{Levine20,Smith17,doi:10.1021/acs.jctc.5b00917,doi:10.1063/1.5140086,doi:10.1063/1.2883976,doi:10.1063/1.2883981}. This active size constraint stymies the application of CASSCF to systems with larger valence spaces, where a small active space is not sufficient and convergence of desired properties with respect to active size is not reached.

This limitation is a potential opportunity for NISQ computers to exhibit a quantum advantage in this keystone method in quantum chemistry, with the active space paradigm often being touted as a near-term prospect for quantum computers \cite{Reiher2017,Takeshita2020} (see e.g. Ref,~\onlinecite{Elfving2020} for a recent review of the limits of classical computers in this field and the requirements for quantum advantage). However, beyond simple analysis of gate depth and qubit number required, is the question of the practical feasibility of a robust and convergent algorithm for the self-consistency of the full CASSCF method, which has to date been recently demonstrated for a single orbital optimization step without full self-consistency in the work of Takeshita {\em et al.} \cite{Takeshita2020}. In the algorithm which we use, the coupling of the active space correlations to the orbital rotations required for self-consistency is provided by the two-body RDM within the active space. Therefore, the faithful sampling of this two-body RDM with sufficient fidelity is critical for a well-behaved algorithm. This is especially important as the orbital rotation parameters involve non-linear functionals of the sampled two-body RDM elements, meaning that we expect noise from the QPU sampling to manifest as systematic error in the final results, even in the case that the sampling of the underlying RDM elements is unbiased. We investigate the two-body RDM active space sampling for this purpose on QPUs as well as the importance of error mitigation, by using a parameterized gate circuit as the active space wave function optimized via the variational quantum eigensolver (VQE). %This is integrated as an iterated subroutine within the fully self-consistent CASSCF algorithm of the PySCF package \cite{https://doi.org/10.1002/wcms.1340,doi:10.1063/5.0006074}.
However, the use of VQE in this work could be replaced by quantum Krylov or imaginary-time solvers suitable for NISQ devices \cite{Wang2019,yeteraydeniz2021benchmarking}, as well as quantum phase estimation algorithms when suitable devices are available.

\subsection{Fully self-consistent algorithm}

We briefly summarize the key steps of the (two-step) CASSCF approach (sometimes also described as the related multi-configurational self-consistent field method), with more details available in Ref.~\onlinecite{Sun2017}. We start with the second quantized electronic Hamiltonian in a basis, as
\begin{equation}
\label{eq:sq_ham}
    \hat{H} = \sum_{ij} h_{ij} \hat{a}_i^{\dagger} \hat{a}_j + \frac{1}{2} \sum_{ijkl} g_{ijkl} \hat{a}_i^{\dagger}\hat{a}_j^{\dagger} \hat{a}_l \hat{a}_k + E_{\mathrm{nuc}},
\end{equation}
where $h_{ij}$ and $g_{ijkl}=\langle ij|kl \rangle$ are the one and two-body integrals respectively, with $E_{\textrm{nuc}}$ the scalar nuclear repulsion.
We parameterize the orbitally-optimized CASSCF wave function of Eq.~\ref{eq:CASSCFWfn} as
\begin{align}
\label{eq:QCAS_WF_1}
    |\Psi_{\mathrm{CASSCF}} \rangle = \ket{\textbf{R}, \textbf{c}} = e^{-\textbf{R}} \sum_n c_n \ket{n},
\end{align}
where $\textbf{R}$ parameterizes the single-particle anti-unitary rotation operator of the molecular orbital basis, and $\textbf{c}$ defines the coefficients of the configurations indexed by $n$, spanning the selected active space. The full optimization problem can then be written as
\begin{align}
\label{eq:QCAS_WF_2}
    E = \min_{\textbf{R}, \textbf{c}} \frac{\bra{\textbf{R}, \textbf{c}}H\ket{\textbf{R}, \textbf{c}}}{\langle{\textbf{R}, \textbf{c}}|\textbf{R}, \textbf{c}\rangle}.
\end{align}

Within the two-step algorithm, the optimization of $\textbf{R}$ and $\textbf{c}$ are treated separately and alternated, as the optimization of $\textbf{R}$ can be efficiently performed for one-body unitary rotations on classical computers, given the knowledge of the active space two-body RDM,
\begin{equation}
    \Gamma_2 = \langle \psi_{\textrm{active}} | \hat{a}_i^{\dagger} \hat{a}_j^{\dagger} \hat{a}_l \hat{a}_k | \psi_{\textrm{active}} \rangle . \label{eq:Gamma2}
\end{equation}
The optimization of ${\textbf R}$ then proceeds via construction of the gradient and Hessian of the energy with respect to these parameters, which can then be updated at modest computational expense via a quasi-second order step, accelerated with iterative subspace methods as implemented in the {\tt PySCF} package \cite{Sun2017,https://doi.org/10.1002/wcms.1340,doi:10.1063/5.0006074}.

For a given rotation matrix parameterized by ${\mathrm R}$, the Hamiltonian can then be transformed as
\begin{align}
\label{eq:QCAS_WF_3}
    H|_\textbf{R} = e^{-\textbf{R}} H e^{-\textbf{R}},
\end{align}
and the Coulomb and exchange contribution from the static core electrons integrated out, resulting in an active space Hamiltonian, $H_{\rm act}({\bf R})$, which only spans the active space degrees of freedom. The optimization of this active space wave function is then amenable to implementation within a VQE minimization, as
\begin{align} \label{eq:CASCI_quantum}
    E|_{\textbf{R}} = \min_{\theta} \bra{\psi_{\textrm{active}}(\theta)} H_{\rm act}({\bf R}) \ket{\psi_{\textrm{active}}(\theta)}, 
\end{align}
where $\theta$ denote the angles to optimize within the chosen quantum circuit parameters \cite{Peruzzo2013}. Once optimized, the 2RDM elements of the active space of Eq.~\ref{eq:Gamma2} can be sampled, in order to update ${\mathbf R}$ in the full space, until convergence. 

\subsection{Results}

While the computational procedures for this coupled orbital optimization are well developed for exact or near-exact solvers in the domain of quantum chemistry, their utility in a fully self-consistent algorithm with a noisy quantum computer is far from clear (although there is some relevant recent work on noisy Monte Carlo solvers for active spaces \cite{doi:10.1063/1.5140086,doi:10.1021/acs.jctc.5b00917,doi:10.1021/acs.jctc.5b01190,doi:10.1080/00268976.2020.1802072}). We therefore consider the CASSCF algorithm with an active space NISQ device solver, to determine the stability of the algorithm in the presence of sampling, gate noise, decoherence, and a parameterized quantum circuit for the state. This allows us to understand the feasibility of this multiscale approach, and develop practical strategies to ameliorate potential shortcomings from the noisy active space sampling.

We apply the method to a carbon monoxide (CO) molecule in a cc-pVDZ basis set, and at a stretched bond length of 1.54~\AA. This stretching of the multiple bond enhances the strong correlation in the electronic structure, as the atomic-like character of the constituent atoms is increased. An active space of two orbitals and two electrons, corresponding to the highest occupied and lowest unoccupied molecular orbitals, is selected to capture the dominant many-body entanglement in these lowest-energy quantum fluctuations. To ensure a significant level of orbital relaxation from the self-consistent procedure, and to test the stability of this noisy optimization in the case of a poor initial choice of orbitals, we select initial orbitals (and active space) from only a partially converged Hartree--Fock calculation. This was achieved by an early stopping of the mean-field self-consistent field procedure after only two updates of the Fock matrix prior to the CASSCF.

\begin{figure*}
    \centering
    \includegraphics[width=\linewidth]{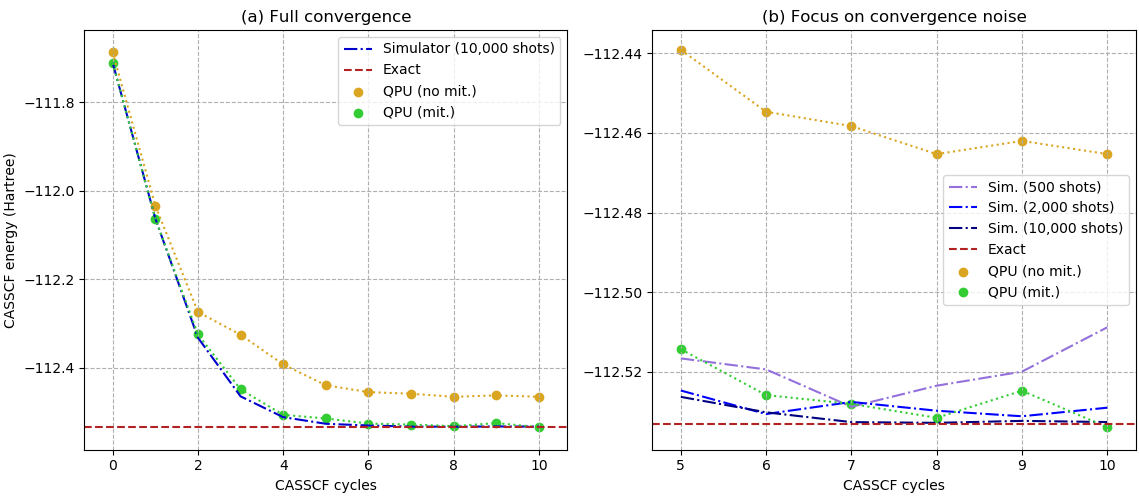}
    \caption{Convergence of the energy of the CASSCF state for a 2-electron, 2-orbital active space of Carbon Monoxide for each orbital update step. Results are shown for a quantum simulator, on IBMQ Bogota (without error mitigation) and on IBMQ Santiago (with error mitigation). Plot (b) focuses on the final five orbital update cycles, showing the variation and bias in converged results, with additional simulated results for 500 and 2,000 shots to illustrate the impact of finite sampling noise (results for 5,000 shots are indistinguishable from those obtained with 10,000 shots and as such were not included).}
    \label{fig:qcasscf_qpu_results}
\end{figure*}

We first implemented the method on a quantum simulator with 500, 2,000, 5,000 and 10,000 shots to sample each mapped two-body RDM operator required, but in the absence of any additional noise model for the gates. For the quantum hardware experiments, we use IBMQ Bogota and IBMQ Santiago, which are both 5-qubit QPUs available through the the IBMQ platform, with equivalent levels of gate fidelity (details about each QPU used in this paper can be found in Appendix \ref{sec:lattices}). The initial calculations on IBMQ Bogota were performed without accounting for any error mitigation, before applying the light-touch error mitigation strategy presented in appendix~\ref{sec:error_mitigation} on the IBMQ Santiago hardware to assess any improvements from this. For the QPU runs, we use a measurement ramp-up schedule whereby the number of measurements is increased if the output energy at a given iteration is higher than for the previous one (which should not be the case during the optimization). It is capped at 8,000 shots, which is also the number of measurements used for RDM sampling after the state is optimized.

We used the same ansatz for all experiments, built on a four-qubit, three-layer version of the HEA \cite{Kandala2017, Mitarai2019}. This resulted in a total of 24 variational parameters in the model. We found it unnecessary to fully converge the ansatz each iteration, and therefore investigated varying the level of ansatz optimization each CASSCF step to improve efficiency. Five iterations of the VQE were in general sufficient on the first cycle, and we then used the parameters obtained to initialise the ansatz for the next cycle. A single iteration of the VQE for subsequent CASSCF steps after performing this warm start was sufficient to fully converge in a reasonable time, and to reach good accuracy. 

The results of these CASSCF optimizations are presented in Fig.~\ref{fig:qcasscf_qpu_results}. Without error mitigation, the QPU results show significant systematic error at convergence of $\sim 60$~\mEh, but nevertheless allow for a stable optimization. Including the error mitigation allows for significantly better results, with fluctuations of less that 10~\mEh~from the exact CASSCF value from exact 2-step optimization of the same initial active space. As expected, the variance from the QPU experiment is significantly more than the corresponding quantum simulated results, even with error mitigation. This reflects the fact that the error mitigation effectively removes the bias in the sampled measurements, but does not materially improve on the variance resulting from quantum noises. In our quantum simulated results, we find strong convergence for any simulation without gate noise or decoherence above 5,000 shots. Below that number, finite sampling noise prevents the algorithm reaching the sought after solution. At 500 shots, it fails to reach under 10~\mEh~difference to the target state energy on average.

We can distinguish and isolate the effects of certain errors arising from the quantum solver on these results. Firstly, we have the systematic error in the VQE at each iteration, including the optimization, gate errors and ansatz choice, which lead to a non-exact energy and state for a given set of active orbitals. Secondly, we can consider the effect of stochastic noise in the RDM due to a finite number of samples. This second error will lead to incorrect orbital updates in the CASSCF macroiterations, and a loss of precision in the final CASSCF energy due to an inability to propagate to the optimal orbitals defining the active space and its Hamiltonian. Furthermore, since the orbital choices in CASSCF are not linear functionals of the sampled density matrix elements, even if the RDM elements are entirely unbiased and correct on average, this does not preclude a systematic error entering the orbital updates at any finite sampling.

\begin{figure}
    \centering
    \includegraphics[width=\linewidth]{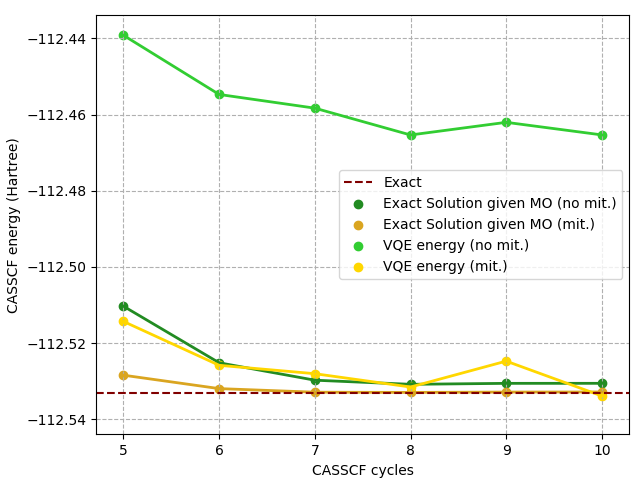}
    \caption{Convergence of the final six orbital update cycles of the CASSCF energy, obtained with the IBMQ Bogota (with no error mitigation) and IBMQ Santiago (with error mitigation). Two series are presented for each QPU calculation: `VQE energy' results are equivalent to Fig.~\ref{fig:qcasscf_qpu_results}, while `Exact solution' represent the exact energy from the current active space, that would have been obtained if the VQE was solved perfectly given the molecular orbitals found from the previous quantum VQE update step.}
    \label{fig:decomp}
\end{figure}

To separate these sources of error, we can consider the exact energy of each CASSCF iteration, but using the active space orbitals obtained at each iteration from the noisy VQE update from the quantum solver. This eliminates errors due to the VQE optimization of a given active space, isolating the error due to convergence of the non-optimal orbital rotations being found at each step, primarily due to the inherent sampling noise of the RDMs. These results are shown in Fig.~\ref{fig:decomp}, and show that the overwhelming majority of the error is arising from the bias in the VQE, while the convergence of the orbitals is highly robust to the errors in the active space VQE description and sampling errors of the RDMs. Even without error mitigation of the RDMs, orbital optimization is accurate to within 10~\mEh, while the results of the simple error mitigation in the RDMs rendered an almost exact CASSCF energy. This demonstrates that the orbital optimization procedure is less susceptible to the errors in the VQE and RDM sampling than the inherent errors in the energy and wave function optimization for a given active space. This relative insensitivity to the fidelty of the RDMs bodes well for larger active space calculations on QPUs and the practicality of orbital optimization through RDM sampling, as well as the improvements which would transfer to this approach from improved active space quantum algorithms \cite{Wang2019,yeteraydeniz2021benchmarking}.

\begin{figure}
    \centering
    \includegraphics[width=\linewidth]{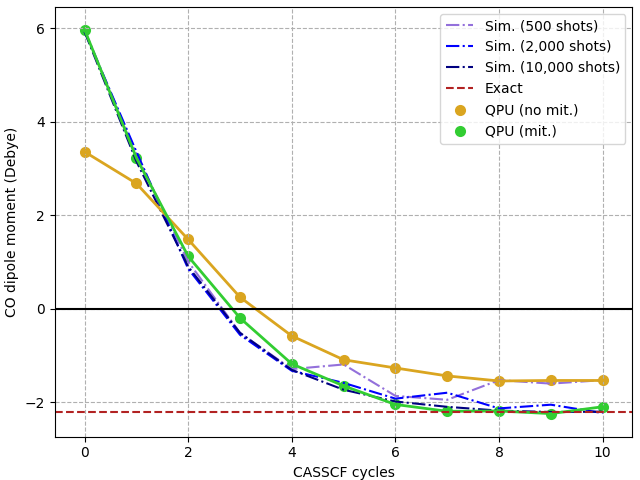}
    \caption{Convergence of the dipole moment in Debye from QPU-CASSCF as the orbitals are optimized, both with (IBMQ Santiago) and without (IBMQ Bogota) error mitigation, as well as quantum simulated results from an RDM sampling of 500, 2,000 and 10,000 shots. Positive dipole moments mean that the dipole moment points towards the oxygen (i.e. the oxygen atom has net negative charge), while the converged results flip the orientation of the dipole moment.}
    \label{fig:dipole}
\end{figure}

A key question remains as to whether this robustness is a property just of orbital optimization, or whether this also extends to a broader set of expectation values which can be derived from the RDM sampling (other than the energy), as these also relax due to a more faithful description of the correlated wave function. We consider here the effect of orbital optimization and a correlated VQE wave function on the magnitude of the dipole moment of the carbon monoxide molecule in the same active space, which can be extracted from the sampled one-body reduced density matrix as a one-particle expectation value. This quantity characterizes the net charge distribution in the molecule, and from symmetry constraints can be described by a vectorial quantity which must be coincident with the carbon-oxygen bond. The magnitude of this vector is shown in Fig.~\ref{fig:dipole}, as the active space orbitals are optimized in the presence of the correlated VQE state.

Since the dipole moments are linear functionals of the one-body RDM, we would expect an unbiased sampling of the RDM to give an unbiased estimate of the dipole moment from the optimized VQE-CASSCF state. We find that without error mitigation, there is still an error of $\sim0.6 D$, however the error mitigated results can effectively reduce the systematic error in the final dipole moment completely, with fluctuations in each cycle of a similar magnitude to the emulated values without quantum noise or decoherence with 5,000 shots. At this point, the fluctuations in the dipole moment agree with the magnitude of the fluctuations expected from the original density matrix sampling experiments of Fig.~\ref{fig:rdm_plot}, and an unbiased estimate of the exact CASSCF dipole moment is obtained. We also note that the correlation and orbital optimization reverses the direction of the charge imbalance in this system from the starting description.

The overall runtime of the full QPU-CASSCF calculations on IBMQ Bogota was $\sim$14 QPU-hours, including 10 orbital updates, VQE optimization and RDM sampling.

However, one must consider the potential for parallelization. In this case, all 8,000 measurements conducted on the 49 operators could have been conducted in parallel, possibly reducing the overall runtime up to well under a second. While this is not a good indication for scaling and long term viability of the method (we encourage the reader to refer to Ref.~\onlinecite{Elfving2020} for an interesting assessment of the scalability of VQE), it does illustrate the potential for strong parallelization, and corresponding error mitigation techniques, for the viability of NISQ algorithms.

Finally, it is worth discussing the viability of extensions to CASSCF on quantum devices. In quantum chemistry, CASSCF is rarely the end of the story, as it neglects the contributions to expectation values arising from interactions between the active space electrons and the external degrees of freedom. These can generally be treated at a perturbative level of theory\cite{Angeli2007}, cumulant or energy-moment expansions \cite{vallury_2020} or via subspace expansions\cite{Takeshita2020}, and are required for quantitative accuracy for predictive calculations. These perturbative couplings between the spaces can be computed by considering the higher-rank RDMs in the active space. This approaches will significantly increase the number of TPB terms which must be sampled. However, large reductions can be found with the appropriate TPB groupings, and this is demonstrated in Appendix~\ref{sec:RDM_group_tables}, where 440,154 Pauli strings for the sampling of the four-body RDM within 6 qubits can be reduced to only 3,182 commuting groups. We will explore the viability of this perturbative extension to CASSCF in future work.

\section{QPU-Enhanced Energy-weighted Density Matrix Embedding} \label{sec:QEwDMET}

The CASSCF method exploits the locality of correlation in the energy domain, choosing and optimizing a low-energy subspace for the correlated treatment. In this section, we demonstrate the utility of a faithful QPU sampling of RDMs in order to correlate and optimize a different subspace, which instead relies on spatial locality. This perspective is often more useful for strongly-correlated extended systems, where the atomic-level correlated degrees of freedom can be isolated, and where widely used methods such as density functional theory fail to provide accurate results \cite{Cohen2012}. These approaches fall under the umbrella of quantum embedding or quantum cluster methods, and are amongst the most promising for QPU-enhanced materials modelling \cite{Maier2005, Kotliar2006, Sun_Embedding16}. We investigate the recently-developed `Energy-weighted Density Matrix Embedding Theory' (EwDMET) as a promising candidate in this direction \cite{Fertitta2018, Fertitta2019,Sriluckshmy2021}.

The EwDMET method connects the density matrix embedding theory (DMET) and dynamical mean-field theory (DMFT), two established approaches in quantum embedding \cite{Knizia2012, Georges1996, Zheng2017}. Both of these `parent' approaches have recently been adapted for use with a quantum hardware solver, as well as related embedding techniques \cite{Bauer2015,Keen2019,Yao2020,dhawan2021dynamical,2019arXiv191004735R,GalliQuantumEmbedding,2016arXiv161006910R,yamazaki2018practical,kawashima2021efficient}. 
However, the EwDMET avoids a number of difficulties in these approaches. Similar to DMET, it avoids any necessity to compute the single-particle Green's function of the resulting quantum cluster problem on the QPU, which is challenging for quantum hardware, although important progress is being made \cite{PRXQuantum.2.010317}.
Instead, the method requires a desired number of one-particle spectral moments from the subspace problem, which can be obtained directly from the reduced density matrices of the ground state. The number of self-consistent spectral moments can then be systematically enlarged, to approach the complete dynamical character of DMFT as an orthogonal polynomial expansion. The method also removes all explicit numerical fitting steps, and constructs a rigorous self-consistency on these spectral moments, systematically extending the DMET formulation and connecting it to its fully dynamical limit. This rigorous and algebraic self-consistency enables non-trivial results to be obtained even at the lowest truncation of the spectral moment expansion. This requirement of only computing ground-state RDMs, while at the same time benefiting from a rigorous and algebraic self-consistency for non-trivial emergent physics makes it an ideal candidate for combination with QPU-derived RDMs in the NISQ era. We briefly review the salient features of EwDMET for this QPU formulation, with more details in Ref.~\onlinecite{Sriluckshmy2021}. 

As with all quantum cluster approaches, the algorithm begins with the choice of a local correlated space. This could be the $d-$shell of a transition metal atom, or a cluster of sites for a discrete lattice model. The EwDMET method then allows for an improvable and self-consistent description of the one-particle quantum fluctuations between this fragment and its environment \cite{Fertitta2018,Fertitta2019,Sriluckshmy2021}. This information is contained within the self-consistently optimized (hole and particle) spectral moments of the fragment, defined as
\begin{align}
T_{h,\alpha\beta}^{(n)} = \bra{\Psi} \hat{a}_{\alpha}^{\dagger} (\hat{H}-E_0)^{n} \hat{a}_{\beta} \ket{\Psi},
\label{eq:15}
\end{align}
\begin{align}
T_{p,\alpha\beta}^{(n)} = \bra{\Psi} \hat{a}_{\alpha} (\hat{H} - E_0)^{n} \hat{a}_{\beta}^{\dagger} \ket{\Psi},
\label{eq:16}
\end{align}
where $\alpha, \beta$ index the degrees of freedom of this local fragment, $n \geq 0$ denotes the order of these moments, optimized up to a maximum desired value \nmom, and $|\Psi\rangle$ is the ground state of the constructed correlated subspace. As \nmom$\rightarrow \infty$, the method exactly reproduces the effective dynamics of DMFT, recast as a ground-state wave function theory, while systematic truncation to low \nmom will still well-describe the dominant low-energy fluctuations from the fragment into its environment. The EwDMET method rigorously maps the full system to a subspace consisting of the chosen fragment coupled to a `bath' space . The size of the bath is determined solely by the size of the fragment and the desired number of spectral moments to capture (and the correlated subspace is hence independent of the size of the full system). It is this correlated subspace problem which must then be solved on the QPU at each iteration, and the spectral moments of Eqs.~\ref{eq:15} and \ref{eq:16} computed. With these computed spectral moments from the correlated fragment space, the one-particle description of the full system can be algebraically updated via the addition of non-interacting auxiliary states, to ensure that the fragment moments at the mean-field level over the full system exactly reproduce the correlated subspace ones. The procedure is iterated, updating the auxiliary space and bath space of the quantum cluster problem, until convergence.

\subsection{Infinitely coordinated Bethe-Hubbard Lattice}

We apply this method to the paradigmatic Hubbard model of condensed matter physics, which describes a range of quantum phases and correlation-driven transitions. Specifically, the limit of an infinitely-coordinated extended Bethe-Hubbard lattice with local interactions defines our model of interest, which has the particular feature that correlation-driven changes to all one-particle properties are site local. This property was used to great effect to motivate the development of DMFT, by providing a non-trivial model for which it describes an exact limit \cite{PhysRevLett.62.324,PhysRevLett.74.809}. The EwDMET has the same exact limit for this model as \nmom$ \rightarrow \infty$.

The model can be equivalently defined in this infinite-dimensional limit via its metallic non-interacting density of states \cite{PhysRevB.71.235119}, which is defined to have the form
\begin{equation}
    A(\omega) = \frac{1}{2 \pi} \sqrt{4-\omega^2} ,
\end{equation}
for a bandwidth of $|\omega| < 2$. This non-interacting spectrum was fit to a single central fragment site with 200 additional degrees of freedom, to approximate this full spectrum to a high energy resolution \cite{Nusspickelwdmft,doi:10.1142/S0217979202011937}. The interacting Hamiltonian is then defined as resulting from the additional on-site Hubbard interaction term, $U {\hat n}_{i \uparrow} {\hat n}_{i \downarrow}$, which is included on the fragment in the correlated subspace Hamiltonian at each iteration. The spectral moments of this central fragment site are then self consistently optimized, where we define the projection of the non-interacting system Hamiltonian into this cluster subspace as $h^{\mathrm clust}$.

In this work, we truncate the spectral moment expansion at order \nmom$=1$, defining the set of self-consistent fragment quantities. This simplifies their computation from the VQE solution for the ground state of the cluster Hamiltonian at each iteration, since these $n=0$ and $n=1$ moments over the fragment can be constructed from (parts of) the one- and two-body RDMs, for which we have efficient sampling as detailed previously. For instance, the $n=1$ hole moment reduces to
\begin{equation}
T_{h,00}^{(1)} =\sum_{j\in\mathrm{clust}}h_{0j}^{\mathrm{clust}}\gamma_{0 j}+U \Gamma_{0000} , \label{eq:rdm_moms}
\end{equation}
where $0$ denotes the fragment site index. Physically, the restriction of \nmom$=1$ means that the center of mass of the particle and hole spectral distributions can be self-consistently optimized on the lattice in the presence of the local correlation effects. This contrasts with DMET, where single site self-consistency cannot change the physics of the full system from the non-interacting picture in translationally symmetric systems, and so QPU emulations of this method with a single fragment site are restricted to single-shot computation without any self-consistency \cite{2016arXiv161006910R,yamazaki2018practical,kawashima2021efficient}. 
Therefore, by using VQE as the solver on quantum hardware, we can identify the ground state of the cluster Hamiltonian, and subsequently sample the relevant RDMs to construct the required fragment spectral moments. 

At this choice of spectral moment truncation, the cluster Hamiltonian consists of the single fragment site and a single bath orbital, resulting in a four-qubit system to solve at each iteration of the EwDMET method. This cluster is solved via VQE on the QPU with the same three-layer HEA as applied in the CASSCF section, with a Jordan-Wigner mapping to the qubit representation. Additionally, the same error mitigation is used to control the noise inherent in the sampling of the RDMs required to construct the fragment spectral moments. Emulated QPU simulations without noise models were also performed for comparison to the QPU experiments of this algorithm.

\subsection{Results}

\begin{figure}
    \centering
    \includegraphics[width=\linewidth]{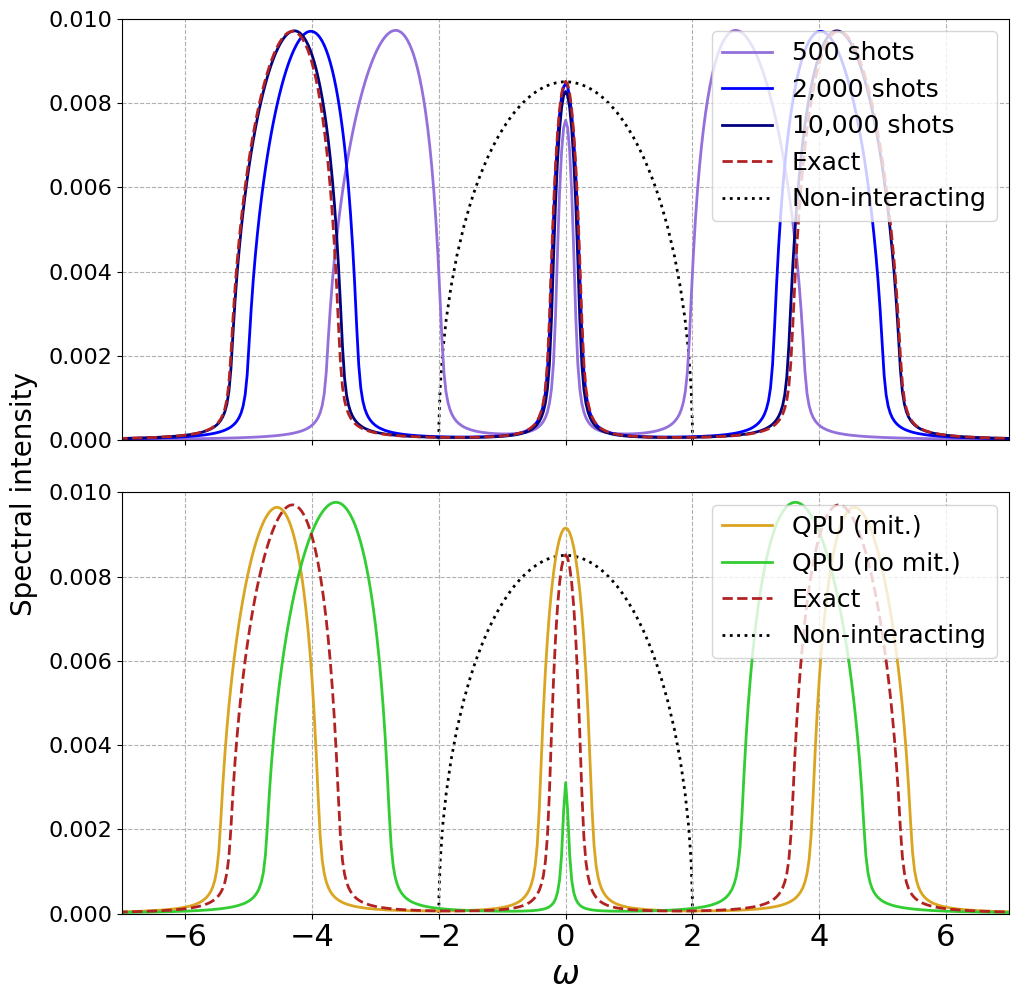}
    \caption{The density of states for the Hubbard model on the Bethe lattice with infinite coordination for the EwDMET method with \nmom$=1$. Upper results are performed on a classical QPU simulator with finite sampling noise and varying numbers of shots in the VQE solution to the cluster Hamiltonian at each iteration. Lower panel shows results from two QPU experiments, on IBMQ Santiago and IBMQ Bogota, with and without error mitigation respectively, with 5,000 shots for the sampling of the required RDMs. Grey dotted lines show the original non-interacting spectrum of the model, while the red dotted line shows the EwDMET(\nmom$=1$) results with an exact solution of the cluster Hamiltonian each iteration.}
    \label{fig:dos}
\end{figure}

Figure~\ref{fig:dos} presents results for the single-particle spectrum for the model at self-consistency for VQE-EwDMET with \nmom$=1$, at a strongly correlated limit with an on-site interaction of $U=8$, which is twice the non-interacting bandwidth of the material. The final converged spectra are presented for an exact solver, an emulated quantum simulation without noise but with different numbers of shots for the sampling of the RDMs (500, 2000 and 10,000), and quantum hardware results on IBMQ machines with and without error mitigation, with 5,000 shots for the sampling. All calculations converge within 8 iterations as the auxiliary and bath spaces adapt to the correlations described over the fragment site in each VQE cluster solution, demonstrating the robustness of the self-consistency in the presence of noise.
At this level of theory, the spectrum shows significant qualitative changes from the non-interacting spectrum. Upper and lower Hubbard bands develop, splitting the original density of states, with a qualitatively correct Mott gap between these bands shown. However, a small quasiparticle peak remains at the Fermi level, showing that the metallic character of the system is not entirely removed by the correlations. Consistency in higher orders of the spectral moments are required to get to a true Mott insulating state \cite{Sriluckshmy2021}. Nevertheless, much of the true correlated spectral density is reproduced at this level of theory. 

At 10,000 shots, the emulated results without further simulated quantum noises are are almost indistinguishable from the exact benchmark at all energies. At lower numbers of shots, the gap between the Hubbard bands is too small, as the variance in the sampled RDMs increases. As with the CASSCF method, the updated auxiliary space at each iteration is a non-linear transformation of the spectral moments (which are themselves linear functionals of the RDM elements, as can be seen in Eqs.~\ref{eq:rdm_moms}). The consequence of this is a systematic error in the resulting spectral functions at convergence due to the increasing RDM variance, rather than simply a manifestation in a noisy but unbiased spectrum. This behavior of an underestimated gap between the Hubbard bands is also present in the QPU results, where unmitigated results feature unrepresentative Hubbard bands. However, the performance is once again considerably improved with the error mitigation, with the Hubbard bands and low-energy peak resolved to higher accuracy, suggesting the method fits well with a QPU cluster solver, and removing the necessity for the full solution of the fragment Green's function at each point within a DMFT framework.

\begin{figure}
    \centering
    \includegraphics[width=\linewidth]{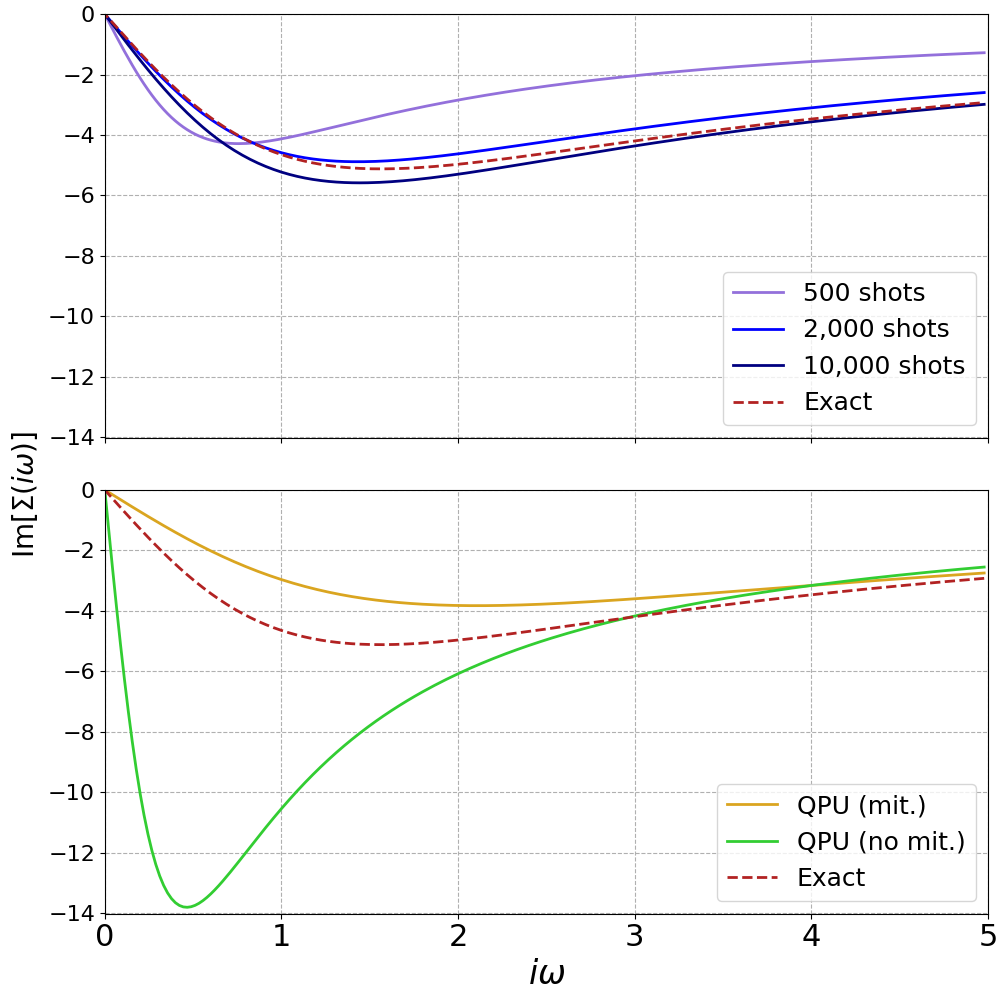}
    \caption{Converged effective on-site Matsubara self-energy for the Bethe lattice Hubbard model with EwDMET at \nmom$=1$. Upper plot shows results from the quantum simulator with finite sampling of the RDM elements, compared to exact results (infinite sampling). Lower plot shows results from quantum hardware on IBMQ Santiago and IBMQ Bogota and 5,000 shots in the RDM sampling, with and without error mitigation in the RDM sampling respectively.}
    \label{fig:self_energy}
\end{figure}

Properties of the system can also be observed from the effective self-energy of the system, which is obtained directly from the self-consistent auxiliary states, and allows access to quantities such as Fermi liquid parameters \cite{Sriluckshmy2021}.
For the same system, Fig.~\ref{fig:self_energy} shows the imaginary part of the self energy on the Matsubara frequency axis. The finite sampling results are seen to approach the exact results with increasing shots, with 10,000 shots reaching a comparable performance to the exact results, with the discrepancy far more visible in the self-energy than the original single-particle spectrum of Fig.~\ref{fig:dos}. The quantum hardware unmitigated results correspondingly demonstrate significant overestimation of the resulting self-energy. Despite the fact that the Hubbard bands are closer, these unmitigated results show a larger effective mass and quasi-particle renormalization at the Fermi surface from the self-energy (a larger derivative at $i\omega \rightarrow 0$), which manifests in the smaller peak in Fig.~\ref{fig:dos} at that point. Error mitigated QPU self-energies are however more in line with exact results for the method, albeit now slightly underestimated at low-frequencies. Further improvement of the results can be obtained by increasing the moment order (\nmom) to which the dynamical quantities are all resolved.  Similar to the perturbative corrections to CASSCF however, these will require sampling of higher-body RDMs, and is an avenue of continuing research.

\section{Conclusions}

We have presented a unified approach to self-consistent coupling of quantum and classical computational resources in quantum chemistry and condensed matter electronic structure problems. This coupling relies on the faithful and efficient sampling of reduced density matrices on quantum resources, where these objects span the correlated physics of an iteratively optimized subspace of the full system. We consider the required fidelity and sampling quality of these density matrices for robust optimization on current generation quantum hardware, as well as their viability for obtaining non-energetic quantities, including the dipole moment for {\em ab initio} simulation of chemical systems, and the self-energy and mass renormalization of strongly correlated extended models. In both of these settings, we also focus on strategies for both reducing the number of terms which are required for sampling these RDM elements, via optimization of a compact tensor product basis, as well as a simple yet effective approaches for error mitigation in the sampled density matrices.

Overall, the picture is encouraging for the viability of these approaches, with the self-consistent optimization found to be particularly robust to the presence of sampling noise on current generation quantum hardware. This self-consistency is found to be more reliable than the uncertainties resulting from the state optimization and energy obtained from the VQE at any single iteration, and points to advantages from the improvements in both hardware and quantum algorithms for state preparation on quantum devices leading to a significant transferal of benefits. These conclusions however are restricted to relatively small correlated subspaces, and further work is required to understand the generality of these conclusions as we access QPU with larger qubit capacity. Furthermore, quantitative rather than qualitative accuracy in these application areas will require a description of higher-body density matrices than sampled in this work. Whilst in this work we begin to investigate the viability of sampling these via a consideration of the compactness of their resulting tensor product basis, this direction will be a key avenue for future research.

\begin{acknowledgments}

J.T. is supported by and industrial CASE (iCASE) studentship, funded by and UK EPSRC [EP/R513143/1], in collaboration with University College London and Rahko Ltd. E.F. is supported by an industrial CASE (iCASE) studentship, funded by the UK EPSRC [EP/T517665/1], in collaboration with the University of Strathclyde, the National Physical Laboratory, and Cambridge Quantum Computing. E.G. is supported by the UK EPSRC [EP/P510270/1].  I.R. acknowledges the support of the UK government department for Business, Energy and Industrial Strategy through the UK national quantum technologies program. G.H.B. gratefully acknowledges support from the Royal Society via a University Research Fellowship, as well as funding from the European Research Council (ERC) under the European Union’s Horizon 2020 research and innovation programme (Grant Agreement No. 759063). This research project is in great part funded by Innovate UK, project Quantifi from UK Research and Innovation, as part of the UK National Quantum Technologies Programme and Industrial Strategy Challenge Fund. We acknowledge the use of IBM Quantum services for this work. The views expressed are those of the authors, and do not reflect the official policy or position of IBM or the IBM Quantum team.

\end{acknowledgments}

\providecommand{\noopsort}[1]{}\providecommand{\singleletter}[1]{#1}%

\newpage
\begin{appendices}

\begin{center}
    \textbf{APPENDICES}
\end{center}

\section{Light touch error mitigation} \label{sec:error_mitigation}

In order to improve the results and reduce the effects of the quantum noises of the IBMQ devices used, we employ an extrapolation approach based on the assumption that the impact of the circuit errors is evenly distributed on the output state (for a more rigorous approach to extrapolation, see Ref.~\onlinecite{Endo2018_PRX}). 
The aim is to recover an approximation of the true expectation value of a quantum circuit with respect to an operator, assuming a specific bias on the output results. It is worth noting that this method primarily focus on correcting gate and readout errors, and as such does not attempt to maintain or restore the purity of the quantum state produced. 
We define the true (desired) expectation value of the circuit with respect to an operator, ${\hat O}$, as $\langle {\hat O} \rangle$, while the measured expectation value is denoted as $\langle$\Omeas$\rangle$.
The outcome of a circuit can be associated with measurement eigenvalues $\pm 1$, with the probability of measuring $1$ equal to $P_1 = \Pr(O_m=1)$, and with $O_m$ referring to a single measurement of operator ${\hat O}$ at the end of the quantum circuit. Similarly, we have $P_1 = 1 - P_{-1} = \Pr(O_m = -1)$. We can associate the true expectation value with:
\begin{equation}
    \langle {\hat O} \rangle = P_1 - P_{-1} = 2P_1 - 1.
\end{equation}

We now assume that there is a certain probability, $P_{err}$, that at least one gate error occurs during propagation and measurement of the quantum circuit. One can assume that there exists a value between $\pm1$, representing the expectation value of ${\hat O}$ given the error rate, which we denote $\langle$\Oerr$\rangle$. This value, as well as the probability distribution of the operator measurements $O_m$ are unknown and cannot be recovered easily.

However, by virtue of the central limit theorem, we can assume that given a sufficiently large number of measurements and a sufficiently deep circuit (with many different errors possible), the random variable $O_m$ will be normally distributed.
This allows us to approximate the expectation value of the measured operator as $\langle$\Omeas$\rangle = (1 - P_{err})\langle {\hat O} \rangle + P_{err} \langle {\hat O}_{err} \rangle$. We can rewrite this as 
\begin{equation}
    \langle {\hat O} \rangle = \frac{\langle {\hat O}_{\textrm meas} \rangle - P_{err} \langle {\hat O}_{err} \rangle}{(1 - P_{err})} .
\end{equation}
We make the further assumption on the value of $\langle {\hat O}_{err} \rangle$ that, given a sufficiently large number of a random circuit errors, the probability of getting either eigenvalue when an error occurs is equal. This method is largely sufficient for the purpose of our experiments, and can act as a lower bound for the benefits error mitigation can achieve with no additional computing cost. With this approximation, and for eigenvalues $\pm 1$, we can ignore this final term, and the expression simplifies to
\begin{equation}
    \langle {\hat O} \rangle = \frac{\langle {\hat O}_{\textrm meas} \rangle}{(1 - P_{err})}, \label{eq:ErrorExtrap}
\end{equation}
while for binary eigenvalues of $1$ and $0$, we get
\begin{equation}
    \langle {\hat O} \rangle = \frac{\langle {\hat O}_{\textrm meas} \rangle - 0.5 P_{err}}{(1 - P_{err})}.
\end{equation}

In order to estimate the final bias on the true value of the operator, we have considered the electron number operator (trace of the one-body RDM). The deviation from the set number of electrons in the system gives us an estimate for the bias induced by quantum noise $P_{err}$. Given our assumptions, the bias factor can be recovered as follows:

\begin{equation}
   \frac{1}{1 - P_{err}} = \frac{N_{\textrm elec}}{N_{\textrm meas}},
\end{equation}
with $N_{\textrm elec}$ the target number of electrons, and $N_{\textrm meas}$ the number of electron measured:
\begin{equation}
   N_{\textrm meas} = \sum_i \gamma_{ii} = \sum_i \bra{\psi} \hat{a}^{\dagger}_i\hat{a}_i \ket{\psi}
\end{equation}

This comes at not extra computational cost as the one-body RDM terms used are necessarily computed as part of the VQE process. 

An alternative method would be to estimate $P_{err}$ directly by computing it through the reported gate calibration data from the QPU provider (compounding the gate fidelities), but we found that in general this approach is less reliable. This is most likely due to the fact that using this latter method treats the bias resulting from circuit errors completely classically: it ignores any part of the bias that could be due to the reduced purity of the quantum state produced, which can otherwise be captured by the former method.  

\section{Operator groupings for RDM element sampling} \label{sec:RDM_group_tables}

\begin{table*}
\begin{ruledtabular}
\begin{tabular}{cccccccccc}
\\       & Active orbitals & 2 & 4 & 6 & 8 & 10 & 12 & 14 & 16 \\\\
\hline\\
     & Fermionic operators   & 3 & 10 & 21 & 36 & 55 & 78 & 105 & 136\\
One-body RDM & JW      & 6 & 28 & 66 & 120 & 190 & 276 & 378 & 496\\

       & JW-TPB & 3 & 9 & 13 & 17 & 21 & 27 & 31 & 36 \\\\
\hline\\
     & Fermionic operators   & 11 & 157 & 786 & 2,486 & 6,085 & 12,651 & 23,492 & 40,156\\
Two-body RDM & JW & 49 & 910 & 4,983 & 16,460 & 41,325 & 87,354 & 164,115 & 282,968\\

       & JW-TPB & 5 & 70 & 227 & 497 & 853 & 1,342 & 1,928 & 2,601 \\\\

\end{tabular}
\end{ruledtabular}
\caption{Number of unique Pauli strings to be measured in the Tensor Product Basis (TPB) in order to sample all elements of the one- and two-body RDMs, as the number of (spatial) orbitals in the active space is enlarged. These numbers represent the terms in a direct Jordan-Wigner mapping (JW) and the Tensor Product Basis based on the grouping of commuting terms from a Largest-degree First Coloring algorithm (JW-TPB).} \label{tab:TPB}
\end{table*}

\begin{table}
\begin{ruledtabular}
\begin{tabular}{cccccccccc}
\\       & Active orbitals & 4 & 6 \\\\
\hline\\
     & Fermionic operators   & 610 & 8,400\\
Three-body RDM & JW & 4,928 & 71,742 \\

       & JW-TPB & 189 & 2,049 \\\\
\hline\\
     & Fermionic operators   & 939 & 40,065\\
Four-body RDM & JW & 11,425 & 440,154\\

       & JW-TPB & 163 & 3,182 \\\\
\end{tabular}
\end{ruledtabular}
\caption{Number of unique Pauli strings to be measured in the Tensor Product Basis (TPB) in order to sample all elements of the three- and four-body RDMs, as the number of (spatial) orbitals in the active space is enlarged. These numbers represent the terms in a direct Jordan-Wigner mapping (JW) and the Tensor Product Basis based on the grouping of commuting terms from a Largest-degree First Coloring algorithm (JW-TPB).} \label{tab:TPB_higher}
\end{table}

In this appendix, we present the approach to operator grouping into the tensor product basis (TPB), as presented in Fig.~\ref{fig:grouping_numbers}. In addition, we also present the TPB for three- and four-body RDMs, up to 6 orbitals, as an investigation into the future feasibility of extended coupling schemes between the quantum region and environment such as multi-reference perturbative approaches. Finding these groups for the higher-body RDMs for larger numbers of orbitals became too computationally demanding for the current algorithm given our resources at time of writing.

In order to find these groupings, we require a graph of commutative relationships between all the Pauli strings required to measure the elements of the RDMs. To find a low number of groups of fully connected sub-graphs, we employed the Largest-Degree First Coloring (LDFC) algorithm (similar to what is proposed in \cite{Hamamura2020}), a graph colouring heuristic. As an example for an alternative to the LDFC algorithm, one can start by grouping Pauli strings according to the frequency of identity operators in the string (as done, for instance in \cite{vallury_2020}).  

The LDFC algorithm works by first grouping Pauli strings which have the highest number of commuting relationships with the other Pauli strings in the set. Once some Pauli strings have been grouped, they are removed from the initial set, and the LDFC algorithm then continues using the same principle. The steps required to complete grouping of Pauli terms (using LDFC), joint measurements and measurement results reconciliation are outlined at a high level below. For a more detailed description, we recommend Ref.~\onlinecite{Gokhale_2019}.

\begin{itemize}
    \item \textbf{Initialization:} From the list of Pauli terms that require grouping, define a graph $G(V, e)$, with $V$ the vertices, corresponding to each Pauli operator, and $e$ the edges representing commuting relationship between Pauli operators.
    \item \textbf{LDFC step 1:} Rank the elements of $V$ according to the number of edges they are connected to. 
    \item \textbf{LDFC step 2:} First allocate a given color to the element of $V$ with the highest rank, and allocate the same color to all elements of $V$ with which it is connected. Continue by allocating a new color to the next element of $V$ in the ranking that is not already allocated to a specific color. 
    \item \textbf{Joint-measurement basis identification:} The groups now been defined. From each group, identify a basis (multiplicative), from which all the other elements of the groups can be computed. The basis size should be $N$, with $N$ the number of qubits. 
    \item \textbf{Joint-measurement circuit construction:} Once a basis is identified, construct the circuit required for joint measurements of the operators (following for instance the instructions set in Ref.~\onlinecite{Gokhale_2019}, aiming to map each of the operators in the basis to a single qubit Z-operator measurement. 
    \item \textbf{Reconciliation:} From the results of the measurement, reconstruct the expectation value of each element in each group that can then be used to compute the one- and two-body RDMs.
\end{itemize}

We present the final number of terms in the TPB for measurement of the RDM elements in Table~\ref{tab:TPB} and \ref{tab:TPB_higher}, arising from the grouping of underlying Pauli strings. These show the substantial reduction in required number of terms to sample. As the LDFC algorithm is a greedy approach to the NP-complete graph colouring problem, it will not guarantee that the resulting TPB is comprised of optimal groups, and as such more efficient grouping could be found. In order to jointly measure all the Pauli strings in a single group, we followed the method presented in Ref.~\onlinecite{Gokhale_2019}.

\section{IBM QPU lattice structures and additional information} \label{sec:lattices}

In this appendix, we present additional information about the IBMQ QPU used during the experiment. The information provided below is sourced from IBMQ  reported calibration of the machines at time of running the experiment and may change slightly overtime. 

\subsection{Lattice structures}

\begin{figure}[H]
    \centering
    \includegraphics[width=\linewidth]{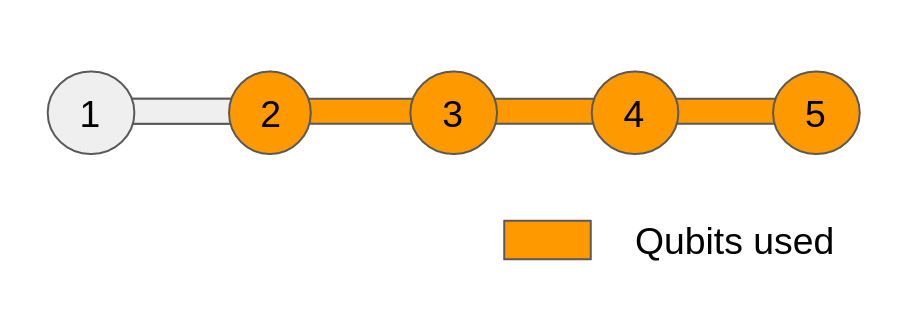}
    \caption{IBMQ Bogota lattice structure and qubits used for experiments. This QPU was used to compute CASSCF and EwDMET with error mitigation.}
    \label{fig:bogota_lat}
\end{figure}

\begin{figure}[H]
    \centering
    \includegraphics[width=\linewidth]{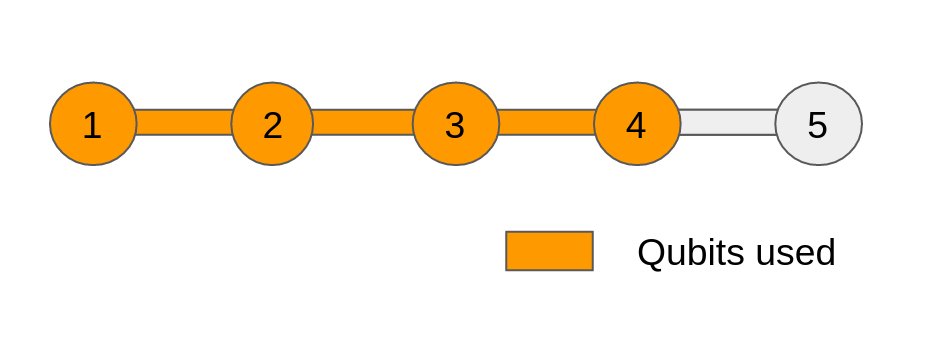}
    \caption{IBMQ Santiago lattice structure and qubits used for experiments. This QPU was used to compute CASSCF and EwDMET without error mitigation.}
    \label{fig:santi_lat}
\end{figure}

\begin{figure}[H]
    \centering
    \includegraphics[width=\linewidth]{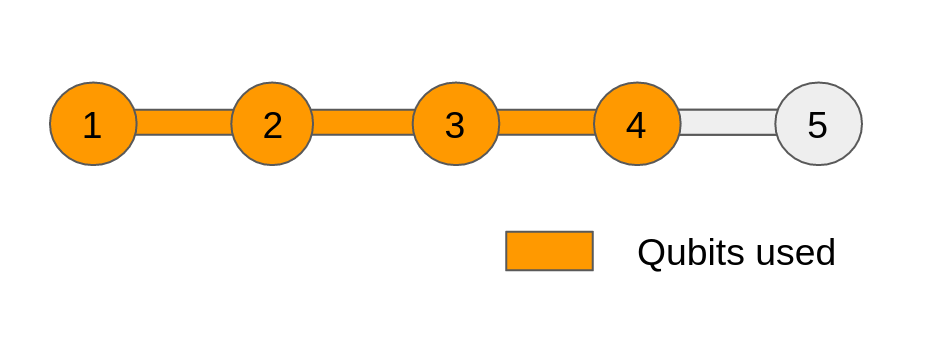}
    \caption{IBMQ Athens lattice structure and qubits used for experiments. This QPU was used to compute the RDM sampling studies presented in Fig. \ref{fig:rdm_plot}, with and without error mitigation}
    \label{fig:athens_lat}
\end{figure}

IBMQ Bogota, Santiago and Athens are all 5-qubit QPUs, following IBM's \textit{Canary} r3 processor type, with reported quantum volume of 32 \cite{Gambetta2020}. The lattice structure, as well as the qubits used are presented in Fig. \ref{fig:bogota_lat}, \ref{fig:santi_lat}, \ref{fig:athens_lat} respectively. 

\subsection{Calibration information}

The information presented in Table \ref{tab:calibration} summarizes the calibration data of the QPU used. It is directly taken from the IBMQ portal and may change overtime as IBM re-calibrates the processors. 

\begin{table}[H]
\begin{ruledtabular}
\begin{tabular}{cccc}
\\ QPU & Bogota & Santiago & Athens \\\\
\hline\\
Single-qubit Pauli-X error & 2.95e-4  & 4.55e-4 & 4.16e-4 \\
Qubit frequency (GHz) & 4.89 & 4.78 & 5.094 \\
Two-qubit gate error & 1.40e-2 & 1.22e-2 & 1.043e-2 \\
Two-qubit Gate time (ns) & 536.89 & 408.89 & 346.67 \\
Read-out length (ns) & 5048.89 & 4017.78 & 3022.22 \\
Read-out error & 3.77e-2 & 1.82e-2 & 1.82e-2 \\
\end{tabular}
\end{ruledtabular}
\caption{Selected calibration metrics from IBMQ. These values are averaged for all qubits / connections and taken at a point in time near the experiment was run. They may change overtime.} \label{tab:calibration}
\end{table}

\end{appendices}

%\appendix
%\begin{widetext}
%\section{Possible appendix}

%\end{widetext}
\end{document}